\documentclass{ar-1col-S2O}
\usepackage{url}
\usepackage[numbers]{natbib}
\usepackage{xspace}
\usepackage{graphicx}
\usepackage{color}
\usepackage{float}
\usepackage{amssymb}
\usepackage{amsmath}
\usepackage{array}
\usepackage{hyperref}
\usepackage{lmodern}
\usepackage{subfigure}
\usepackage{slashed}
\usepackage{mathrsfs}
\renewcommand{\theenumi}{(\arabic{enumi})}
\usepackage{enumitem}
\def\DDbar{D^0-\overline D{}^0}

\def\beq{\begin{equation}}
\def\eeq{\end{equation}}

\newcommand{\bea}{\begin{eqnarray}}
\newcommand{\eea}{\end{eqnarray}}

\jname{Annu. Rev. Nucl. Part. Sci.}
\jvol{76:In press}
\jyear{2026}
\doi{10.1146/annurev-nucl-100324-023932}

\begin{document}

\markboth{Petrov and Zheng}{Super Tau Charm Factories}

\title{The Physics and Prospects of Super-Tau Charm Factories}

\author{Alexey A. Petrov$^1$ and Yangheng Zheng$^2$
\affil{$^1$Department of Physics and Astronomy, University of South Carolina, Columbia, SC, USA, 29208; email: apetrov@sc.edu}
\affil{$^2$School of Physical Sciences, University of the Chinese Academy of Sciences, Beijing, China, 100049}
}

\begin{abstract}
The proposed Super tau-charm factories are a powerful new class of high-luminosity electron-positron colliders operating in the center-of-mass energy range between 2 and 7 GeV, a region that spans thresholds for tau leptons, open-charm hadrons, charmonium and charmonium-like states, hyperons, and light hadrons. With unprecedented data samples, threshold kinematics, and quantum-coherent production, these facilities offer unique opportunities to advance precision tests of the Standard Model and to search for physics beyond it. In this review, we examine the physics prospects of the Super Tau-Charm Facility, focusing on precision charm measurements, CP violation in mesons and baryons, tau-lepton properties and rare decays, and nonperturbative QCD phenomena such as hadronization, spectroscopy, and time-like form factors. We also discuss the experimental landscape, technological challenges, and complementarity with existing and planned facilities. Together, these capabilities position super tau-charm factories at the forefront of the precision frontier in particle physics.
\end{abstract}

\begin{keywords}
tau-charm factory, charmed mesons, QCD, tau physics, CP-violation
\end{keywords}
\maketitle

\tableofcontents

\section{Introduction}
\label{Section1}

The Super Tau--Charm Facility (STCF) is envisioned as a next-generation electron--positron collider operating at center-of-mass energies from approximately 2 to 7~GeV, with a design luminosity exceeding $5\times10^{34}\,\mathrm{cm^{-2}s^{-1}}$ and upgrade potential to $10^{35}\,\mathrm{cm^{-2}s^{-1}}$. This energy range spans a uniquely rich and underexplored region of the Standard Model (SM), covering thresholds for tau-lepton production, open-charm hadrons, charmonium and charmonium-like states, hyperon--antihyperon pairs, and a wide variety of light hadronic final states. With extremely large data samples of processes featuring threshold kinematics and quantum coherence, in a low-background experimental environment, STCF is well positioned to address several fundamental open questions in particle physics.

The physics program of the proposed STCF is driven by three overarching scientific themes: the nature of color confinement and hadron formation; the origin and manifestation of symmetry violations, such as CP-violation and the search for physics beyond the Standard Model (BSM)  through precision measurements and rare processes \cite{Petrov:2021idw}. These themes are interconnected and uniquely accessible in the tau--charm energy regime, which lies at the transition between perturbative and nonperturbative QCD \cite{Brambilla2014,Alkofer2011}. STCF possesses excellent capabilities for studying exclusive reactions, threshold behavior, and quantum-entangled final states, thereby enabling qualitatively new experimental approaches.

A defining strength of STCF is its ability to produce unprecedentedly large, clean samples of tau leptons, charmed and light-quark hadrons. Over a nominal decade of operation, annual integrated luminosities at the inverse-attobarn level are anticipated, yielding samples of order $10^{10}$ tau pairs and comparably large yields of charmed mesons, charmed baryons, and hyperons. These statistics are competitive with, and in several channels complementary to, those of Belle II, while offering the decisive advantage of operating at or near production thresholds~\cite{Kou2019}. Threshold operation provides powerful kinematic constraints, suppresses combinatorial backgrounds, and enables quantum-correlation techniques that would otherwise be inaccessible.

Charm physics is a major pillar of the STCF program. The facility's energy range spans thresholds for $D\bar D$, $D_s\bar D_s$, and charmed-baryon production, enabling precision measurements of decay constants, form factors, and absolute branching fractions using quantum-correlated initial states~\cite{Bigi2012}. Measurements of charm mixing and CP violation benefit directly from the coherent production of $D^0\bar D^0$ pairs. Rare charm decays are sensitive to flavor-changing neutral currents that are suppressed in the Standard Model \cite{Petrov:2021idw}. Above 5~GeV, the production of excited charmonium-like states and open-charm resonances provides essential input for establishing reliable charmed-hadron spectroscopy and for disentangling conventional states from exotic configurations~\cite{Chen2016}.

Studying hyperons at STCF offers another avenue for testing CP symmetry. Billions of quantum-entangled hyperon--antihyperon pairs produced in $J/\psi$ decays provide access to CP-violating observables with sensitivities approaching $10^{-4}$ and potentially $10^{-5}$ with polarized beams. These measurements probe CP violation in the baryon sector at a level comparable to Standard Model expectations and offer discovery potential for new CP-violating sources. The same datasets enable measurements of hyperon magnetic and electric dipole moments, thereby placing stringent constraints on CP-violating interactions \cite{HeMa2023}.

In the tau sector, STCF enables a comprehensive program of precision tests of lepton universality, measurements of fundamental tau properties, and searches for charged-lepton flavor violation. Threshold scans enable precise determinations of the tau mass, while lifetime measurements benefit from reduced boosts and excellent vertexing. The large tau samples provide sensitivity to rare and forbidden decays at the $10^{-9}$--$10^{-10}$ level, probing energy scales far beyond those directly accessible at colliders \cite{Petrov:2021idw,Calibbi2018}. These measurements place the tau lepton at the center of STCF's new-physics reach, complementing muon-based precision experiments and high-energy searches.

A particularly distinctive feature of STCF is its role as a factory for exotic hadrons. The tau--charm region is densely populated with the so-called $XYZ$ states, whose internal structure remains one of the central puzzles in contemporary hadron physics \cite{Esposito2017}. STCF's ability to scan energies with high precision, together with large statistics and excellent detector resolution, enables mapping of line shapes, quantum numbers, and decay patterns. In addition, the enormous production rates of $J/\psi$ and other charmonium states enable detailed analyses and searches for gluonic excitations and glueball-like states \cite{Olsen2015}.

STCF provides a unified experimental environment for the simultaneous study of hadronization, spectroscopy, flavor dynamics, and symmetry tests. Precision measurements of hadronic cross sections and electromagnetic form factors are directly relevant to determining the hadronic contribution to the muon anomalous magnetic moment~\cite{Davier2020}. By exploiting exclusive final states and multidimensional observables, STCF also serves as a testing ground for modern QCD tools.

In this review, we discuss the STCF's role at the precision frontier of high-energy physics. Section \ref{Section2} examines the physics prospects of modern tau-charm factories. Section \ref{Section3} covers the current experimental landscape. Section \ref{Section4} discusses the experimental design and technological challenges of the STCF. We conclude in Section 5.

\section{Physics Prospects of Super-Tau Charm Factories}
\label{Section2}
\subsection{Precision Charm Physics}

Charm physics plays a central role in the STCF's precision program, where threshold production, quantum coherence, and very large data samples of charmed states combine to enable measurements of unprecedented accuracy. By employing the double‑tag technique—reconstructing both charm hadrons produced at threshold—STCF will provide a uniquely clean experimental environment for studying charm decays, mixing, and CP violation. It will enable precise determinations of absolute branching fractions, decay constants, and form factors with minimal model dependence, while also providing powerful constraints on strong phases and hadronic amplitudes. STCF's precision charm program will also test the Standard Model in a sector where theoretical control is improving rapidly, allowing us to glimpse new dynamics at energy scales far beyond those directly accessible at colliders.

\subsubsection{Leptonic and semileptonic charm decays}

Leptonic and semileptonic decays of charmed mesons and baryons are among the most precise probes of weak interactions and hadronic structure, and they provide tests of the Standard Model. Precision studies of charmed-hadron weak decays yield measurements of the Cabibbo--Kobayashi--Maskawa (CKM) elements $|V_{cs}|$ and $|V_{cd}|$, test lepton-flavor universality (LFU), and probe non-standard interactions. Recent theoretical progress, driven by lattice QCD and improved continuum methods, such as QCD sum rules, and rigorous dispersive parameterizations of form factors, has reduced many hadronic uncertainties to the percent level in several channels, preparing the field for experimental progress at STCF \cite{STCF:PhyDetCDR}.

In the Standard Model, the tree-level decay width for the leptonic decay $D_q^+\to\ell^+\nu_\ell$  is
\begin{equation}\label{eq:leptonic-width}
\Gamma(D_q^+\to\ell^+\nu_\ell)
=\frac{G_F^2}{8\pi}\,|V_{cq}|^2\,f_{D_q}^2\,m_\ell^2\,m_{D_q}\left(1-\frac{m_\ell^2}{m_{D_q}^2}\right)^2,
\end{equation}
where $m_\ell$ is the charged-lepton mass and $V_{cq}$ denotes $V_{cd}$ or $V_{cs}$ for $q=d,s$ as appropriate. The decay is helicity suppressed by $m_\ell^2$, which makes the muonic and especially the electronic modes rare compared to $\tau$ modes when kinematically allowed. Here, the hadronic matrix element of the axial current is parameterized by the decay constant $f_{D_q}$:
\begin{equation}
\langle 0|\bar q\gamma^\mu\gamma_5 c|D_q(p)\rangle = i f_{D_q} p^\mu .
\end{equation}
Radiative and electroweak corrections modify Eq.~\eqref{eq:leptonic-width} at the few-per-mille to percent level. Structure-dependent contributions, such as internal bremsstrahlung or structure-dependent emission, are important for experimental acceptance and for high-precision theory-experiment comparisons. 

Semileptonic decays factorize into leptonic and hadronic currents. For a heavy pseudoscalar meson $D_q$ decaying to a pseudoscalar $P=\pi,K,\eta$ one has the vector current decomposition
\begin{align}\label{eq:DtoP-decomp}
\langle P(p')|\,\bar q\gamma^\mu c\,|D_q(p)\rangle
&= f_+(q^2)\!\left[(p+p')^\mu-\frac{m_D^2-m_P^2}{q^2}q^\mu\right] + f_0(q^2)\frac{m_D^2-m_P^2}{q^2}q^\mu,
\end{align}
with $q^\mu=(p-p')^\mu$ and the kinematic constraint $f_+(0)=f_0(0)$. Neglecting the charged-lepton mass, which is an excellent approximation for $e$ and a very good one for $\mu$ except near kinematic endpoints, the differential decay rate is
\begin{equation}\label{eq:diff-rate-DP}
\frac{d\Gamma(D\to P\ell\nu)}{dq^2}
= \frac{G_F^2\,|V_{cq}|^2}{24\pi^3}\,|{\bf p}'|^3\,|f_+(q^2)|^2,
\end{equation}
where $|{\bf p}'|=\lambda^{1/2}(m_D^2,m_P^2,q^2)/(2m_D)$ and $\lambda(a,b,c)=a^2+b^2+c^2-2ab-2ac-2bc$. For $\ell=\tau$ or the BSM scalar currents, $f_0(q^2)$ contributes with coefficients proportional to $m_\ell^2/q^2$.

The Standard Model predicts universal $V-A$ couplings for all three generations of leptons, so the only difference among $e$, $\mu$, and $\tau$ flavored states would be in the available phase space, as reflected in the lepton mass factors in Eqs.~(\ref{eq:leptonic-width}) and (\ref{eq:diff-rate-DP}). This implies that searches for possible NP-induced non-$V-A$ or generation-dependent interactions are feasible at STCF. In particular, ratios like 
\begin{equation}
R_{P}^{\ell_1/\ell_2}(q^2)\equiv\frac{d\Gamma(D\to P\ell_1\nu)/dq^2}{d\Gamma(D\to P\ell_2\nu)/dq^2}
\end{equation}
are predicted with high accuracy in the SM, including phase-space effects and radiative corrections. Their comparison with experiment constrains lepton-flavor non-universal interactions. Both total rates and differential ratios provide complementary sensitivity.

Non-$(V-A)$ interactions can also be probed in semileptonic decays by examining deviations in rates, shapes, and angular observables. A general low-energy effective Hamiltonian for $c\to q\ell\nu$ transitions can include scalar and tensor operators,
\begin{equation}
\mathcal{L}_{\rm eff}=-\frac{4G_F}{\sqrt2}V_{cq}\Big[(1+g_V)\,\bar q\gamma^\mu P_L c\,\bar\ell\gamma_\mu P_L\nu + g_S\,\bar q\,c\,\bar\ell P_L\nu + g_T\,\bar q\sigma^{\mu\nu}c\,\bar\ell\sigma_{\mu\nu}P_L\nu + \ldots\Big],
\end{equation}
where $g_{S,T}$ parameterize NP couplings normalized to the Fermi constant. Scalar currents mainly modify contributions proportional to $f_0(q^2)$ (or longitudinal/helicity-suppressed amplitudes), while tensor currents affect angular observables. Precision studies of the form factors are therefore crucial for establishing tight bounds on $g_{S,T}$.

\subsubsection{Rare decays with lepton flavor conservation}

The ultimate goal of low-energy BSM physics studies is to precisely determine the Wilson coefficients of the Standard Model Effective Field Theory \cite{Petrov:2021idw,Grzadkowski:2010es}, which encapsulate all heavy BSM degrees of freedom. The SM EFT Lagrangian could then be matched to a low-energy effective Lagrangian describing rare charm transitions governed by the $c \to u \ell^+ \ell^-$ current, 
\beq\label{SeriesOfOperators2}
{\cal L}_{\rm BSM}^{\rm rare}  = 
- \frac{1}{\Lambda^2} \sum_{i=1}^{10}  \widetilde C_i (\mu) ~ \widetilde Q_i,
\eeq
where ${\rm \widetilde C}_i$ are Wilson coefficients, $ \widetilde Q_i$ are the effective operators, and $\Lambda$ denotes the energy scale of BSM interactions that generate the $ \widetilde Q_i$ operators. There are only ten such operators with canonical dimension six. The first five are listed as
\bea\label{SetOfOperatorsLL}
\begin{array}{l}
\widetilde Q_1 = (\overline{\ell}_L \gamma_\mu \ell_L^\prime) 
(\overline{u}_L \gamma^\mu
c_L)\ , \quad 
\widetilde Q_2 = (\overline{\ell}_L \gamma_\mu \ell_L^\prime)  
(\overline{u}_R \gamma^\mu
c_R)\ , \quad 
\widetilde Q_3 = (\overline{\ell}_L \ell_R^\prime) \ (\overline{u}_R c_L) \ , \\
\widetilde Q_4 = (\overline{\ell}_R \ell_L^\prime) 
(\overline{u}_R c_L) \ , \qquad \quad
\widetilde Q_5 = (\overline{\ell}_R \sigma_{\mu\nu} \ell_L^\prime) 
( \overline{u}_R \sigma^{\mu\nu} c_L)\ , 
\end{array}
\eea
and five additional operators $\widetilde Q_6, \dots, \widetilde Q_{10}$ obtained from those in Eq.~(\ref{SetOfOperatorsLL}) by interchanging $L \leftrightarrow R$. For $\ell^\prime = \ell$, the effective Lagrangian in Eq.~(\ref{SeriesOfOperators2}) is general and includes the SM contribution. The Wilson coefficients in Eq.~(\ref{SeriesOfOperators2}) can be computed using conventional methods \cite{Petrov:2021idw}. Other rare decays, such as radiative $D \to \rho \gamma$ transitions, receive dominant SM contributions that are difficult to compute.
 
The simplest rare decay is a purely leptonic transition of a neutral $D$-meson into a lepton pair, $D^0 \to \ell^+ \ell^-$. Its branching fraction can be written as 
\beq\label{llgenb}
{\cal B} (D^0 \to \ell^+\ell^-) = 
\frac{M_D}{8 \pi \Gamma_{\rm D}} \sqrt{1-\frac{4 m_\ell^2}{M_D^2}}
\
\left[ \left(1-\frac{4 m_\ell^2}{M_D^2}\right)\left|A\right|^2  +
\left|B\right|^2 \right],
\eeq
where $A$ and $B$ are the (complex) amplitudes that depend on the short-distance Wilson coefficients in Eq.~(\ref{SeriesOfOperators2}) and on long-distance parameters. It is essential to note that all non-perturbative QCD effects in that transition can be parameterized by a single $D$-meson decay constant, which can be calculated using lattice QCD or other nonperturbative methods,
\beq \label{DlCoeffb}
\left| A\right| = \frac{f_D M_D^2}{4 \Lambda^2 m_c} \left[\widetilde C_{3-8} + 
\widetilde C_{4-9}\right],  \ 
\left| B\right|  = \frac{f_D}{4 \Lambda^2} \left[
2 m_\ell \left(\widetilde C_{1-2} + \widetilde C_{6-7}\right)
\right. 
+ \left. \frac{M_D^2}{m_c}
\left(\widetilde C_{4-3} + \widetilde C_{9-8}\right)\right]\,, 
\eeq
with $\widetilde C_{i-k} \equiv \widetilde C_i-\widetilde C_k$. Note that matrix elements of certain operators in Eq.~(\ref{SetOfOperatorsLL}), or of their linear combinations, vanish due to vector current conservation and other arguments \cite{Petrov:2024ujw}. 

\begin{figure}[t]
\begin{center}
\includegraphics[width=6cm]{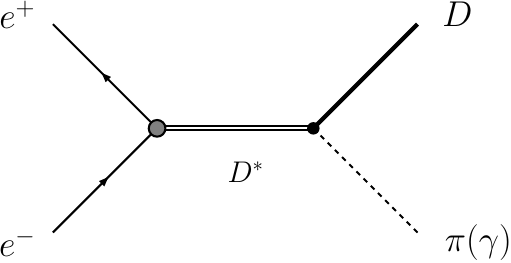}
\end{center}
\caption{\label{DirectRare} Probing the $ c\bar{u}\to e^+ e^-$ vertex
with the $D^*(2007)^0$ production in $e^+e^-$ collisions (from \cite{Khodjamirian:2015dda}).}
\end{figure}

We note that, due to helicity suppression, tests of lepton universality, i.e., the lepton flavor dependence of the Wilson coefficients in Eq.~(\ref{DlCoeffb}), in $D^0 \to \mu^+ \mu^-$ versus $D^0 \to e^+ e^-$ are experimentally very challenging. The most stringent constraints on these decays are given by \cite{HeavyFlavorAveragingGroupHFLAV:2024ctg},
\beq\label{DeeDmumu}
{\cal B}(D^0 \to \mu^+\mu^-) < 3.1 \times 10^{-9}, 
\
{\cal B}(D^0 \to e^+ e^-) < 7.9 \times 10^{-8}, 
\eeq
An interesting alternative to studying $c \to u e^+ e^-$ in leptonic $D$ decays is to measure the corresponding {\it production} process $e^+e^- \to D^*(2007)$. This process, shown in Fig.~\ref{DirectRare}, was proposed in \cite{Khodjamirian:2015dda}. It is feasible if STCF is tuned to operate at the center-of-mass energy equal to the $D^*$ meson mass, $\sqrt{s} \approx 2007$ MeV. This energy region could be covered at STCF while measuring $e^+e^- \to$ hadrons for $R(s)$.

The technique searches for the produced $D^{*0}$ resonance, tagged by a single charmed particle in the final state, that decays strongly ($D^{*0} \to D^0\pi^0$) or electromagnetically ($D^{*0}\to D^0\gamma$) with branching fractions of $(61.9\pm 2.9)\%$ and $(38.1\pm 2.9)\%$, respectively. Although this process is very rare, it offers clear advantages for BSM studies compared to the $D^0 \to e^+e^-$ decay: helicity suppression is absent, and a broader set of effective operators can be examined. Additionally, unlike other rare decays of charmed mesons, long-distance SM contributions are well understood theoretically and contribute at the same order of magnitude as short-distance effects. Finally, one gains a phase space factor (typically, $16\pi^2$) due to the fact that the process $e^+e^- \to D^*(2007)$ is a $2\to 1$ production process, compared to the $2\to 2$ D-meson production in $e^+ e^- \to D^{(*)} \bar D^{(*)}$. There is an additional $1/\alpha$ gain since the $e^+ e^- \to D^{(*)} \bar D^{(*)}$ production proceeds through electromagnetic interactions. 

The CMD-3 Collaboration performed the first studies of this process \cite{Shemyakin:2020uye}, which resulted in an upper limit reported as a branching ratio for the decays of the $D^{0*}(2007)$,
\beq
{\cal B}(D^{0*}(2007) \to e^+ e^-) < 1.7 \times 10^{-6}. 
\eeq
Alternatively, the decays $D^{0*}(2007) \to \mu^+\mu^-$ (or $e^+ e^-$) can be searched for by analyzing $B^- \to \pi^- \mu^+\mu^-$ decays with $B^- \to \pi^- D^{0*} (\to \mu^+\mu^-)$. The LHCb collaboration reported an upper limit \cite{LHCb:2023fuw},
\beq
{\cal B}(D^{0*}(2007) \to \mu^+\mu^-) < 2.6 \times 10^{-8}. 
\eeq
Studying the BSM contributions to rare charm decays can help probe correlations among processes, such as $\DDbar$ mixing and rare decays \cite{Golowich:2009ii}. Usually, it is not possible to predict the rate of a rare decay from the mixing rate alone, even when both the mixing and ${\cal B}(D^0 \to \ell^+\ell^-)$ are dominated by a single operator. However, this is possible for a limited set of BSM models \cite{Golowich:2009ii,Fajfer:2023nmz}. 

Similar searches can be performed for the rare decays when $\ell^\prime \neq \ell$. Experimental limits on ${\cal B}(D^0 \to \mu^+e^-)$ \cite{LHCb:2015pce}
\beq\label{Demu}
{\cal B}(D^0 \to \mu^+e^-) < 1.3 \times 10^{-8}, 
\eeq
give constraints on lepton-flavor-violating interactions. Similar limits can also be obtained from two-body charmed quarkonium decays~\cite{Hazard:2016fnc}.

\subsubsection{Charm meson mixing and CP violation}

Neutral charm meson mixing and CP violation provide a uniquely sensitive probe of flavor dynamics in the up-type quark sector. In the Standard Model, both effects are highly suppressed by the Glashow--Iliopoulos--Maiani mechanism and by the smallness of down-type quark mass splittings, making theoretical predictions challenging and experimental measurements especially valuable. As a result, charm mixing and CP violation are widely regarded as promising windows into physics beyond the Standard Model, complementary to studies in the kaon and $B$-meson systems. The effective Hamiltonian ${\cal H} = M - i\Gamma/2$, with $M$ and $\Gamma$ being the mass and decay matrices,  governs the time evolution of the neutral $D$ system. Off-diagonal pieces in that Hamiltonian mix the $D^0$ and $\bar D^0$ states, resulting in the new propagating mass eigenstates with differing masses and lifetimes \cite{Petrov:2024ujw}. Such mixing is conventionally characterized by the dimensionless parameters
\beq
x \equiv \frac{\Delta m}{\Gamma}, \qquad
y \equiv \frac{\Delta \Gamma}{2\Gamma},
\eeq
with $\Delta m$ and $\Delta\Gamma$ the mass and width differences between the mass eigenstates and $\Gamma$ the average decay width. In the Standard Model, both $x$ and $y$ are expected to be at or below the percent level, with long-distance contributions dominating theoretical uncertainties \cite{Petrov:2024ujw,Friday:2025gpj,Lenz:2020awd}. The SM relies on intricate cancellations among the $d$, $s$, and $b$ quark contributions, making it an excellent arena for testing our understanding of the interplay of weak and strong interactions. Experimental measurements of the $\DDbar$ mixing parameters place strong constraints on theoretical models and potential new physics effects. Although recent progress at LHCb, Belle II, and other experiments has yielded precise measurements of the mixing parameters, the theoretical understanding of the SM contribution to $\DDbar$ mixing remains a challenge  \cite{Golowich:1998pz,Falk:2001hx,Cheng:2024hdo,Bigi:2000wn,Dulibic:2025emg}. Recently, lattice QCD approaches have been developed to address this problem \cite{DiCarlo:2025mnm}. 

The Super Tau--Charm Facility offers decisive experimental advantages for studying charm mixing. At the $\psi(3770)$ resonance, $D^0\bar D^0$ pairs are produced in a coherent quantum state with well-defined charge conjugation. This coherence enables time-integrated methods that are unavailable in uncorrelated production environments. In particular, measurements of double-tagged decay rates provide direct access to mixing parameters and strong-phase differences without relying on time-dependent decay reconstruction. The decay rate for correlated decays into final states $f_1$ and $f_2$ can be written schematically as
\beq
\Gamma(f_1,f_2) \propto 
\left| A_{f_1}\bar A_{f_2} - \bar A_{f_1} A_{f_2} \right|^2,
\eeq
where $A_f$ and $\bar A_f$ denote the decay amplitudes of $D^0$ and $\bar D^0$. These relations enable precise determinations of the strong phases, which are essential inputs for global charm and other flavor analyses. Table~\ref{tab:STCF-mixing} summarizes representative STCF statistical sensitivities from fast-simulation studies assuming 1~ab$^{-1}$ samples collected at several energy points (4009, 4015, and 4030~MeV). The studies consider multiple final states and both $C$-even quantum-correlated and incoherent samples, such as $D \bar D \gamma$. The values shown are the projected one-standard-deviation uncertainties. The most precise results are obtained at 4030~MeV, where larger cross sections and available channels maximize sensitivity \cite{STCF-report}.
\begin{table}
\centering
\caption{Representative STCF statistical sensitivities for $D^0$--$\bar D^0$ mixing and CPV parameters from fast-simulation studies assuming 1~ab$^{-1}$ per energy point \cite{STCF-report}.}
\label{tab:STCF-mixing}
\begin{tabular}{lcccc}
\hline\hline
Mode & Energy (MeV) & $\delta x, \times10^{-4}$ & $\delta y, \times10^{-4}$ & $\delta\phi$ \\
\hline
$K^-\pi^+\pi^+\pi^-$ & 4009 & $4.7$ & $2.5$ & $3.10^\circ$ \\
                     & 4015 & $4.3$ & $2.1$ & $2.85^\circ$ \\
                     & 4030 & $4.3$ & $2.1$ & $2.83^\circ$ \\
$K_S\pi\pi$         & 4009 & $6.9$ & $5.0$ & $4.57^\circ$ \\
                     & 4015 & $6.4$ & $4.6$ & $4.24^\circ$ \\
                     & 4030 & $6.3$ & $4.6$ & $4.19^\circ$ \\
$K^-\pi^+\pi^0$     & 4009 & $4.4$ & $1.7$ & $2.51^\circ$ \\
                     & 4015 & $4.1$ & $1.6$ & $2.34^\circ$ \\
                     & 4030 & $4.0$ & $1.6$ & $2.25^\circ$ \\
\hline\hline
\end{tabular}
\end{table}

Comparison with other programs highlights the complementarity of STCF. Belle~II and LHCb aim for excellent sensitivities using large boosts or very large samples at higher energies. Belle~II projections reach competitive precision in certain multi-body Dalitz analyses, and LHCb provides very small statistical uncertainties in prompt and $B$-decay-tagged samples \cite{Kou2019}. The unique power of STCF lies in its threshold/coherent dataset and its ability to determine strong phases and absolute branching fractions with minimal model dependence. Combined fits that incorporate STCF inputs will therefore significantly improve global determinations of mixing and CP-violating parameters.

\subsection{CP-Violation Studies at the Super Tau--Charm Facility}

\subsubsection{Theoretical Background}

CP violation in the Standard Model originates from the irreducible phase in the CKM matrix. In the up-type sector, these effects are strongly suppressed because the relevant amplitudes involve products of small CKM elements and are generated predominantly through higher-order processes. In the SM, the magnitude of quark-sector CP violation may be expressed by the Jarlskog invariant \cite{Jarlskog},
\beq
J = \mathrm{Im}\!\left[V_{ij}V_{kl}V_{il}^{*}V_{kj}^{*}\right] \sim 3\times 10^{-5}.
\eeq
As a consequence, the areas of all CKM unitarity triangles are equal and proportional to $J$, which makes charm transitions a clean window into suppressed SM phases and possible new-physics contributions \cite{Petrov:2021idw,Bigi:1999hr}. Extraction of CKM matrix elements enables precise testing of the CKM unitarity relations, such as the second-row unitarity relation
\begin{equation}
|V_{cd}|^2+|V_{cs}|^2+|V_{cb}|^2 \stackrel{?}{=} 1.
\end{equation}
Combining leptonic and semileptonic D-decay results allows determinations of $|V_{cd}|$ and $|V_{cs}|$ with competitive precision. At current precision levels, uncertainties from radiative corrections, isospin breaking, and lattice systematics must be included in the error budget.

CP violation in charmed hadron transitions can be seen through the decay asymmetry
\beq
A_{CP}(f) = \frac{\Gamma(D\rightarrow f) - \Gamma(\bar{D}\rightarrow \bar{f})}{\Gamma(D\rightarrow f) + \Gamma(\bar{D}\rightarrow \bar{f})},
\eeq
where $f$ is the final state. Such asymmetry can be either time-dependent (CP-violation in D-mixing) or time-independent (direct CP-violation). In the SM, it is expected to be at most a few parts in $10^{-3}$, making it highly sensitive to new phases.

In the lepton sector, the SM contribution to CP violation in $\tau$ decays is extremely small. Effects could arise from exotic scalar, vector, or leptoquark interactions, and are conveniently parameterized in Low Energy Effective Field Theory (LEFT). For processes involving electromagnetic interactions, the CP-odd observable of interest is the electric dipole moment of the tau,
\beq
\mathcal{L}_{\mathrm{EDM}} = -\frac{i}{2} d_{\tau} \bar{\tau}\sigma^{\mu\nu}\gamma_{5}\tau F_{\mu\nu},
\eeq
whose SM prediction, $d_\tau \sim 10^{-37}$ e $\cdot$ cm, is unobtainably small \cite{Yamaguchi:2020eub}. Consequently, any nonzero EDM of the $\tau$-lepton within foreseeable sensitivity would signal new physics.

\subsubsection{Experimental Advantages of STCF}

Although a full introduction to the facility is provided in Section \ref{Section3}, specific features of STCF are crucial to its CP-violation program. Operations at charm and $\tau$ thresholds ensure quantum-coherent production of $D^{0}\bar{D}^{0}$ and $\tau^{+}\tau^{-}$ pairs, enabling analyses inaccessible at higher-energy machines. At the $D\bar{D}$ threshold, the initial state is produced as a pure $C=-1$ configuration, allowing extraction of the relative strong phases through interference among various decay modes and providing unique pathways to study CP-violation \cite{Qcorrelations}. Such measurements underpin all precision determinations of charm mixing and CP-violating parameters.

The clean environment associated with threshold running substantially reduces combinatorial backgrounds, enabling the analysis of rare decay channels and time-integrated or time-dependent CP asymmetries with high purity. Vertex resolution, particle identification, and hermetic calorimetry together enable full event reconstruction, which is particularly important for semileptonic channels and invisible final-state signatures. The projected integrated luminosities yield event samples of approximately $10^{10}$ charm pairs and comparable numbers of $\tau^{+}\tau^{-}$ events, significantly reducing statistical limitations compared with previous charm factories.

\subsubsection{CP Violation in Charm Decays}

CP violation in the neutral charm system can arise from three sources: CP violation in decay, in mixing, and in the interference between mixing and decay \cite{Petrov:2024ujw}. These effects may be parameterized through
\beq
\left|\frac{q}{p}\right| \neq 1, \qquad
\phi \equiv \arg\!\left(\frac{q\bar A_f}{p A_f}\right) \neq 0,
\eeq
where $p$ and $q$ relate flavor and mass eigenstates,
\beq
|D_{1,2}\rangle = p |D^{0}\rangle \pm q |\bar{D}^{0}\rangle,
\eeq
satisfies $|q/p|\neq 1$ or if the interference phase $\phi = \arg(q\bar{A}_{f}/(pA_{f}))$ differs from zero. The SM expectation for $\phi$ is below the $10^{-3}$ level, implying that any sizable measurement of CP-violating phases or asymmetries would indicate physics beyond the SM. 

STCF’s large, clean charm samples enable precise measurements of time-integrated CP asymmetries and mixing-induced observables across a wide range of decay modes, including CP eigenstates, semileptonic decays, and multibody hadronic final states. The coherent production environment significantly reduces systematic uncertainties from detector asymmetries and decay-time resolution while enabling consistency checks across multiple channels. Sensitivities at or below the $10^{-4}$ level are expected for several key observables, extending into the range where SM effects and plausible new-physics contributions may compete, with modes such as $D^{0}\to K^{+}K^{-}$ and $D^{0}\to\pi^{+}\pi^{-}$ serve as benchmark channels \cite{Petrov:2024ujw}. Since the initial $D\bar D$ states are produced in the quantum-coherent state,
\beq\label{QCoh}
|\Psi\rangle = \frac{1}{\sqrt{2}}(|D^{0}\rangle|\bar{D}^{0}\rangle - |\bar{D}^{0}\rangle|D^{0}\rangle),
\eeq
new measurements are possible \cite{Petrov:2021idw,Qcorrelations}. With quantum-correlated data, strong phase information becomes experimentally accessible via double-tagged events, thereby improving not only asymmetry measurements but also amplitude analyses in multibody decay modes. Dalitz-plot analyses can be pursued in both model-dependent and model-independent forms, the latter relying on binning schemes that reduce sensitivity to poorly known scalar-resonance behavior.

Indirect CP violation is probed through lifetime ratios and effective decay widths. The quantities $A_{\Gamma}$ and $y_{CP}$, defined by
\beq
A_{\Gamma} = \frac{\tau(D^{0}\to f) - \tau(\bar{D}^{0}\to f)}{\tau(D^{0}\to f) + \tau(\bar{D}^{0}\to f)},
\qquad
y_{CP} = \frac{\tau(K^{-}\pi^{+})}{\tau(h^{+}h^{-})} - 1,
\eeq
are especially sensitive to deviations in $|q/p|$ or $\phi$. STCF's threshold environment enables measurement of effective lifetimes without ambiguity in production flavor, thereby reducing systematic uncertainties from tagging and detector-induced asymmetries. Projected sensitivities at the $10^{-4}$ level \cite{STCF-report} will enable indirect CP violation to be probed well below existing constraints.

\subsubsection{Quantum Coherence and CPT Tests}

The quantum coherence of the $D^{0}\bar{D}^{0}$ system at threshold opens a distinct avenue to study fundamental symmetries. The antisymmetric initial state of Eq.~(\ref{QCoh}) produces correlations between decay times and final states that allow sensitive tests of decoherence phenomena \cite{STCF-report}. 

CPT violation may be examined using a formalism akin to that developed for the kaon system. Introducing a small CPT-violating parameter into the effective Hamiltonian governing $D^{0}$--$\bar{D}^{0}$ mixing alters both the mass and width differences and generates distinctive patterns in decay-time distributions and correlated decay rates. Threshold data allow clean access to these effects because the production flavor is fixed and backgrounds are minimal. 

The large $J/\psi$ samples and clean $J/\psi \to \Lambda\bar\Lambda$ topologies planned for STCF create an exceptional laboratory for baryon-sector CP tests. Quantum-entangled hypero-antihyperon pairs enable measurements of asymmetries in angular distributions and decay parameters, with projected sensitivities that can approach or surpass Standard Model expectations, especially when beam polarization is available. Moreover, STCF can perform systematic studies of hyperon magnetic and electric dipole moments by exploiting spin-correlation observables in threshold production. These baryonic probes open a complementary window onto sources of CP and T violation and provide stringent constraints on models that generate dipole-type operators \cite{STCF-report}.

\subsubsection{CP Violation in the Tau Sector}

Producing $\tau^{+}\tau^{-}$ pairs near threshold enables precise control of spin correlations, which are essential for measuring CP-odd combinations of angular observables. The EDM induces characteristic changes in the differential cross section and in the angular distributions of $\tau$ decay products. By exploiting well-defined initial-state quantum numbers and reconstructed decay kinematics, STCF can extract the real and imaginary components of $d_{\tau}$ with high precision. Expected sensitivities reach the $10^{-18}\,e\cdot\mathrm{cm}$ level, an order-of-magnitude improvement over current bounds \cite{TauEDM}.

Hadronic $\tau$ decays, such as $\tau^{\pm}\to K_{S}\pi^{\pm}\nu_{\tau}$, provide additional tests. These modes exhibit inherent CP asymmetries from $K^{0}$--$\bar{K}^{0}$ mixing, predicted to yield an asymmetry of about $3.6\times10^{-3}$ in the SM \cite{TauKS}. Precise measurement of this asymmetry at STCF serves both to validate experimental techniques and to probe potential non-SM enhancements. More complex hadronic final states with three or more pseudoscalars allow tests of CP-odd triple-product correlations, which are sensitive to BSM interactions. 

\subsection{Tau Lepton Physics}

The tau lepton provides a rich laboratory for precision tests of the Standard Model and for indirect searches for physics beyond it. Its relatively large mass, short lifetime, and broad decay spectrum provide sensitivity to electroweak couplings, dipole moments, and flavor-violating mechanisms inaccessible in the lighter-lepton sectors. A high-luminosity $e^+e^-$ collider operating in the tau--charm region enables a program of precision measurements and rare-decay searches complementary to efforts at B factories and hadron colliders. In such an environment, threshold kinematics, clean final states, and high-performance tracking and calorimetry collectively enable significant advances in determining the tau mass and lifetime, measuring its magnetic moment and other electromagnetic form factors, and exploring lepton-flavor-violating decays across a broad range of final states. Finally, studies of quantum mechanics, such as measuring entanglement and Bell nonlocality in the $\tau^+\tau^-$ state near and above its kinematic threshold, are possible at STCF \cite{Han:2025ewp}.

A central objective of tau physics is to improve the precision of the tau mass $m_\tau$, which enters numerous observables, from leptonic decay widths to electroweak universality tests. The current world-average uncertainty of roughly 0.12 MeV \cite{PDG2024} leaves room for substantial refinement. Near threshold, the cross section for $e^+e^- \to \tau^+\tau^-$ depends sharply on $m_\tau$. The theoretical threshold shape is well understood, with the tree-level form
\beq
\sigma(s)= \frac{4\pi\alpha^2}{3s}\,\beta\!\left(1 - \frac{\beta^2}{3}\right),
\eeq
modified by initial- and final-state radiation, vacuum polarization, and Coulomb interactions \cite{KuraevFadin,Jadach2000}. Here $\beta = \sqrt{1 - 4m_\tau^2/s}$ and $s$ is the squared center-of-mass energy. With precise beam-energy determination and fine-grained threshold scans, the mass can be extracted with a sensitivity of several $\times 10$ keV. The tau leptonic decay width
\beq
\Gamma(\tau^- \rightarrow \ell^- \bar\nu_\ell \nu_\tau)
= \frac{G_F^2 m_\tau^5}{192\pi^3}\, f\!\left(\frac{m_\ell^2}{m_\tau^2}\right)\!\left[1+\delta_{\mathrm{RC}}\right],
\eeq
depends on $m_\tau$ with radiative corrections $\delta_{\mathrm{RC}}$ known to high accuracy \cite{MarcianoSirlin,Arbuzov2006}. Improved measurements of $\Gamma(\tau^- \rightarrow \ell^- \bar\nu_\ell \nu_\tau)$ therefore strengthen determinations of observables sensitive to LFU. Near threshold, the small tau momentum yields decay lengths that are measurable with excellent vertexing and well-characterized interaction-point profiles. With anticipated detector resolutions at the few $\times 10~\mu$m level, a sub-femtosecond determination of $\tau_\tau$ appears attainable. Such accuracy would meaningfully affect the results of universality tests.

The tau's electromagnetic properties, particularly its anomalous magnetic moment $a_\tau = (g_\tau - 2)/2$, remain among the least well-constrained parameters in lepton physics. Standard Model predictions, dominated by QED contributions with smaller electroweak and hadronic corrections, are known to better than $10^{-7}$, whereas direct experimental bounds are still two to four orders of magnitude weaker \cite{PDG2024}. Given the expected $m_\ell^2/\Lambda^2$ scaling of many BSM operators, the tau's large mass enhances sensitivity to heavy states coupling to charged leptons \cite{Bernabeu2007,FaelPassera2019}.

A high-luminosity threshold facility offers several promising approaches for measuring $a_\tau$. Radiative tau-pair production, $e^+e^- \to \tau^+ \tau^- \gamma$, probes the electromagnetic vertex via distortions in the photon energy and angular spectra. More generally, the full tau electromagnetic vertex
\beq
\Gamma^\mu(q^2) = F_1(q^2)\gamma^\mu + i\frac{F_2(q^2)}{2m_\tau}\sigma^{\mu\nu} q_\nu,
\eeq
can be constrained by fitting multidimensional distributions of tau decay products, exploiting the strong spin correlations at threshold. The form factor $F_2(0)$ determines the anomalous magnetic moment, while $F_2(q^2>0)$ encodes possible deviations from pointlike behavior or contributions from higher-dimensional operators. A comprehensive program combining radiative production, spin-correlation observables, and global effective-field-theory frameworks may plausibly reach sensitivities of $10^{-4}$ or lower for $a_\tau$.

Searches for charged-lepton flavor violation in tau decays constitute another major component of the physics program. In the Standard Model extended only by neutrino masses, LFV branching fractions are minuscule, often below $10^{-50}$, so any experimental signal would be clear evidence of BSM physics \cite{Calibbi2018,Hisano1996}. A statistical sample of $10^{10}$--$10^{11}$ tau pairs enables sensitivity to LFV branching fractions at or below $10^{-9}$, with certain modes potentially probing the $10^{-10}$ regime depending on the detector \cite{STCF-report}. Radiative modes such as $\tau \to \mu\gamma$ and $\tau \to e\gamma$ offer clean kinematic signatures. Purely leptonic modes, including $\tau \to \ell_1 \ell_2 \ell_3$, achieve strong background suppression through invariant-mass and missing-energy constraints. Semileptonic channels, such as $\tau \to \ell P^0$ with $P^0 = \pi^0, \eta, \eta'$, and baryonic final states, including $\tau \to \Lambda \ell$, are also well suited for reconstruction in a threshold environment. 

The combination of hermetic detector coverage, high particle-identification performance, and manageable QED backgrounds is particularly advantageous for exclusive LFV searches. The resulting sensitivity would probe or exceed predictions from some of the most compelling new-physics frameworks currently under consideration. By exploiting the unique advantages of tau production near threshold, STCF can significantly advance the frontier of lepton physics.

\subsection{QCD and hadronic Physics}

The STCF provides a uniquely powerful experimental environment for studying QCD in the energy regime where confinement, hadronization, and strong-coupling dynamics dominate. Operating at center-of-mass energies from approximately 2 to 7~GeV with high luminosity and fine energy tunability, STCF spans thresholds for light hadrons, baryon--antibaryon pairs, open-charm mesons and baryons, and a dense spectrum of charmonium and charmonium-like states. This energy domain lies at the transition between perturbative and nonperturbative QCD, where theoretical descriptions are least constrained and experimental guidance is most urgently needed.

A central objective of the hadronic physics program at STCF is to elucidate the mechanism of color confinement and the emergence of hadrons from quarks and gluons. Electron--positron annihilation provides a clean initial state with well-defined quantum numbers, enabling systematic studies of hadronization without complications from initial-state hadrons. For instance, meson fragmentation functions are well constrained at high energies but remain poorly understood in the few-GeV regime. STCF enables systematic measurements of hadron multiplicities, correlations, and energy-flow observables across a wide range of energies, providing direct constraints on fragmentation and hadron formation in the confinement-dominated region.

Hadron spectroscopy is another major pillar of the STCF hadronic program. The tau--charm energy region features a rich resonance spectrum, including conventional quark--antiquark states, excited charm hadrons, and a large number of exotic candidates. In particular, the proliferation of charmonium-like $XYZ$ states near open-charm thresholds has raised fundamental questions about the organization of the hadron spectrum beyond the naive quark model~\cite{Olsen2015,Esposito2017}. STCF's ability to perform fine-grained energy scans with high luminosity is ideally suited to disentangling genuine resonant behavior from kinematic effects associated with coupled channels and threshold singularities. Precision measurements of masses, widths, line shapes, and decay angular distributions across many final states will enable stringent tests of competing interpretations of exotic states, including hadronic molecules, compact multiquark configurations, and cusp phenomena~\cite{Chen2016}. Above 5~GeV, STCF will also produce large samples of excited charmonium and open-charm states, enabling the construction of a reliable charm-hadron spectroscopy, while shedding light on how confinement and chiral symmetry breaking manifest in systems containing heavy quarks.

Baryon physics is another area where STCF offers exceptional opportunities. Near-threshold production of baryon--antibaryon pairs probes strong final-state interactions and spin-dependent dynamics in a clean environment. In particular, the enormous samples of $J/\psi$ decays anticipated at STCF will yield billions of quantum-entangled hyperon--antihyperon pairs. The joint angular distribution of baryon decays,
\beq
\frac{d^2\Gamma}{d\Omega_1\,d\Omega_2}
\propto
1+\alpha_1\alpha_2\,\hat n_1\!\cdot\!\hat n_2,
\eeq
encodes information on decay parameters, polarization, and spin correlations, providing sensitive probes of nonperturbative QCD dynamics in baryons. Also, STCF will provide a platform for studying the electromagnetic structure of hadrons in the time-like region. 


\subsection{Light New Physics}

High-luminosity $e^+e^-$ flavor factories, like STCF, provide a great chance to search for rare processes where $D$ mesons decay into final states that leave no traces in the detector or invisible final states. Such invisible final states could indicate feebly interacting light new-physics particles, such as dark photons or axion-like particles (ALPs). Depending on their masses and couplings to the SM particles, they may or may not be candidates for dark matter. If these particles decay after leaving the detector, their experimental signature (or the absence of one) resembles that of neutrinos. These searches require high-purity final states, which can be achieved at STCF. 

Thus, the only unavoidable SM background with the same experimental signature is heavy meson decays into final states containing only neutrinos. The branching ratio for a $D$-meson to decay into a pair of neutrinos is proportional to a very small factor $x_\nu^2= m^2_\nu/M^2_D$,
\beq\label{Dto2nu}
{\cal B}(D\rightarrow \nu\overline{\nu})= \frac{G_F^2\alpha^2f_{D}^2 M_{D}^3} {16 \pi^3 \sin^4
\theta_W\Gamma_{D}} \left|\sum_{k} \lambda_k X^l(x_k)\right|^2 x_\nu^2.
\eeq
For neutrino masses $m_\nu \sim \sum_i m_{\nu_i} < 1$~eV, where $m_{\nu_i}$ is the mass of a neutrino, Eq.~(\ref{Dto2nu}) predicts the branching ratio ${\cal B}_{th}(D^0 \to \nu\bar\nu) \simeq 1\times10^{-30}$, which is unobservable \cite{Badin:2010uh}. 

Thus, transitions to invisible final states constitute almost background-free modes for searches for new light particles. To finish the argument, we should note that, experimentally, the $\nu \bar \nu$ final state does not serve as a good representation of the invisible width of $D^0$ or $B^0$ mesons in the Standard Model. In fact, in the SM, the final state that cannot be detected at a collider includes an arbitrary number of neutrino pairs \cite{Bhattacharya:2018msv},
$
\mathcal{B}\left(D\rightarrow\slashed{E}\right)=\mathcal{B}\left(D\rightarrow\nu\bar{\nu}\right)
+\mathcal{B}\left(D\rightarrow\nu\bar{\nu}\nu\bar{\nu}\right)+\dots .
$
As Eq.~(\ref{Dto2nu}) shows, the decay to the $\nu\bar \nu$ final state is helicity-suppressed.The four-neutrino final state, however, is not.  Thus, it is expected to have a considerably larger branching ratio,
\beq\label{Ratio}
\frac{\mathcal{B}\left(D\rightarrow\nu\bar{\nu}\nu\bar{\nu}\right)}{\mathcal{B}\left(D\rightarrow\nu\bar{\nu}
\right)} \sim \frac{G_F^2 M_D^4}{16 \pi^2 x_{\nu}^{2}} \gg 1.
\eeq
Still, a calculation \cite{Bhattacharya:2018msv} shows that the SM prediction for the invisible width of a heavy meson remains small,
${\cal B} (D^0 \to \nu\bar \nu \nu\bar \nu) = (2.96 \pm 0.39) \times 10^{-27}$ \cite{Bhattacharya:2018msv}. The Belle collaboration has searched for ${\cal B}(D^0 \to \slashed{E})$ \cite{HeavyFlavorAveragingGroupHFLAV:2024ctg}, 
\beq\label{Demu}
{\cal B}(D^0 \to \slashed{E}) < 9.4 \times 10^{-5},
\eeq
which provides constraints on the couplings of the invisible light states. Branching fractions for heavy meson states decaying into $\chi_s \overline\chi_s$ and $\chi_s \overline\chi_s\gamma$, where $\chi_s$ is a dark matter particle of spin $s$, can be computed within the EFT framework. Because the production of scalar $\chi_0$ states avoids helicity suppression, the decay of a $D^0$ state into a final state containing a pair of $\chi_s$ particles can provide constraints on their properties.

\section{Experimental Landscape}
\label{Section3}
\subsection{Overview of past and current tau-charm factories (BEPC-II, CLEO-c)} 

The tau–charm energy region is central to particle physics, and several dedicated experiments have long explored this regime. These facilities operate $e^+e^-$ colliders at center-of-mass energies in the tau–charm range and record the resulting collision products for detailed analysis. Tau–charm factories provide clean data samples, low backgrounds, excellent detection efficiency and resolution, and a well-defined initial state (known quantum numbers and four‑momentum). An additional advantage is pair production at threshold and the associated quantum‑correlation effects, which enable unique precision measurements.
   
Historically, many dedicated experiments have made major contributions, including the three generations of MARK experiments at SPEAR (USA), the DM2 experiment at the DCI collider in Orsay (France), the BES and BESII experiments at the BEPC in Beijing, and CLEO‑c at Cornell (USA).
   
The BEPCII/BESIII program is a major upgrade of the original BEPC/BESII facility. Construction began in 2005 and data taking started in 2009. Its current center‑of‑mass energy range covers 2–5 GeV (with a planned upgrade to 5.6 GeV), and its peak luminosity has reached $1.1 \times 10^{33} \, \mathrm{cm}^{-2} \, \mathrm{s}^{-1}$. BESIII remains the only operating tau–charm facility worldwide and, since 2009, has exploited capabilities such as energy scans, charmonium resonance production, and threshold measurements to deliver important results in light‑hadron spectroscopy, charmonium and exotic states, nucleon structure, charmed‑hadron decays, CP‑violation studies, and searches for new physics.
   
Compared with the design luminosity of the next generation $e^+e^-$ B factory SuperKEKB/Belle II \cite{Belle II:Acc, Belle II:Det}, BESIII's peak luminosity is approximately two orders of magnitude lower, so its data volume and detector performance will no longer meet future physics goals. In 2024 BEPCII completed an upgrade (BEPCII‑U) that can increase luminosity by a factor of three in the 4–5 GeV range. However, infrastructure constraints make any increase in order of magnitude in BEPCII infeasible, creating an urgent need for a new generation ultrahigh-luminosity tau-charm facility.
   
Accordingly, the Super Tau‑Charm Facility (STCF) has been proposed in China \cite{HEPChina:2016} and the Super Charm-Tau factory (SCTF) has been proposed in Russia \cite{SCT, SCTF:Exp}. To meet physics requirements, the accelerator must deliver a center‑of‑mass energy range of 2 to 7 GeV with continuous tunability; achieve a luminosity above $0.5 \times 10^{35} \, \mathrm{cm}^{-2} \, \mathrm{s}^{-1}$ at the optimized energy near 4 GeV and maintain stable long‑term operation; and enable a detector spectrometer capable of efficient and precise measurements under conditions of high radiation, high background, high event rates, and wide dynamic range.

\subsection{Advancements in accelerator technology and luminosity goals}
    
   The key collider parameters are the center-of-mass energy and luminosity. For symmetric colliders, the center-of-mass energy is twice the beam energy. In principle, collider luminosity scales positively with beam energy, is inversely proportional to the vertical beam envelope (beta) function at the interaction point (IP), and is proportional to the beam-beam parameter in the vertical plane. To achieve higher luminosity, the beams must be extremely strongly focused at the IP while suppressing beam instabilities that grow with beam current, and the crossing angle and collision point must be precisely controlled. Consequently, the design of a new-generation collider requires precise analysis of beam dynamics, and its technical systems demand $\text{R\&D}$ and innovation in areas such as ultra-strong magnetic fields, highly stable RF systems, and beam diagnostics with high spatiotemporal resolution.
   
   The development and construction of electron–positron colliders worldwide have spanned over 50 years. The first generation of circular $e^+e^-$ colliders adopted a single-ring scheme; representative facilities include ADONE at Frascati in Italy, SPEAR at SLAC in the United States, and BEPC in China. In single-ring colliders, the positron and electron beams coexist around the full ring in non-collision sections and interact with each other; moreover, the number of bunches, beam current, collision frequency, and the performance of various technical systems are tightly constrained, resulting in relatively low luminosity. For example, BEPC achieved a luminosity of $1 \times 10^{31} \, \mathrm{cm}^{-2} \, \mathrm{s}^{-1}$. To overcome the drawbacks of first-generation colliders, the second generation adopted a dual-storage-ring configuration and, through multi-bunch operation, achieved ampere-level effective beam currents, dramatically increasing luminosity. In the same energy range, second-generation $e^+e^-$ colliders improved luminosity by about two orders of magnitude over the first generation. The first group of second-generation double-ring colliders was completed in the late 1990s: DA$\Phi$NE in Italy (1997), reaching $5 \times 10^{32} \, \mathrm{cm}^{-2} \, \mathrm{s}^{-1}$; KEKB in Japan (1999) and PEP-II in the United States (1999), among which KEKB, in the 10–12 GeV center-of-mass energy range, long held the world record luminosity of $10^{34} \, \mathrm{cm}^{-2} \, \mathrm{s}^{-1}$. China’s BEPCII, completed in 2009, is also a second-generation circular $e^+e^-$ collider; in 2016 its luminosity reached $1 \times 10^{33} \, \mathrm{cm}^{-2} \, \mathrm{s}^{-1}$, 100 times that of BEPC.
   
   As research has advanced, experiments have required data with much higher statistics, and a new generation of $e^+e^-$ colliders must substantially boost luminosity even further. However, because of limitations from collective effects and power, pushing beam current up by additional orders of magnitude has hit a bottleneck. Over the past two decades, after extensive study, the international accelerator-physics community has reached a consensus: for new-generation circular $e^+e^-$ colliders, the primary route to higher luminosity is ultra-strong focusing in the interaction region, compressing the vertical beta function at the IP to the sub-millimeter scale; at the same time, to effectively mitigate the hourglass effect, a large crossing angle is adopted; and to suppress the beam-beam resonances and associated luminosity loss induced by the large crossing angle, a pair of crab-waist sextupole magnets is installed in the IR, effectively raising the luminosity. To cope with the strong nonlinearities introduced by ultra-strong focusing and the beam instabilities associated with very high beam currents, next-generation collider designs must comprehensively analyze the complex, multidimensional, coupled beam dynamics in the interaction region to obtain optimal solutions. In 2007 INFN proposed the large crossing angle (large Piwinski angle) with Crab Waist scheme, which was experimentally validated in 2009 \cite{PhiFactory}. Since 2010, this scheme has become a common strategy for the design of high-luminosity circular $e^+e^-$ colliders and has been implemented in the SuperB factory \cite{SuperB:Acc}, the SCTF projects in Russia \cite{SCT} and Italy \cite{SuperB:SCTF}, and the CEPC \cite{CEPC:AccTDR} design in China. It is anticipated that the luminosity of third-generation $e^+e^-$ colliders will exceed that of second-generation machines in the same energy range by more than a factor of 50. The first third-generation circular $e^+e^-$ collider, SuperKEKB, began commissioning in 2015. It originally adopted a design combining a large Piwinski angle with the nano‑beam scheme, but at the end of 2019 the lattice was redesigned to incorporate a crab‑waist scheme, which was installed thereafter. During 2024 operations it achieved a luminosity of $4.47 \times 10^{34} \, \mathrm{cm}^{-2} \, \mathrm{s}^{-1}$, already higher than that of the second-generation B factories, but still about an order of magnitude below its design goal of $6 \times 10^{35} \, \mathrm{cm}^{-2} \, \mathrm{s}^{-1}$. Continued accelerator-physics studies and technological $\text{R\&D}$ for third-generation circular $e^+e^-$ colliders are therefore required. STCF will adopt the large-Piwinski angle with Crab Waist scheme and the luminosity goal is to exceed $0.5 \times 10^{35} \, \mathrm{cm}^{-2} \, \mathrm{s}^{-1}$ \cite{STCF:AccCDR} .
    
\subsection{Tau-charm physics reach of other experiments (LHCb, Belle II)}

 Another representative $e^+e^-$ collider program relevant to tau-charm physics is the B factory. Historically, B factories include the BaBar experiment at SLAC in the United States and the Belle experiment at KEK in Japan. The Belle experiment at KEK (the High Energy Accelerator Research Organization in Japan) operated on the KEKB $e^+e^-$ collider with a center-of-mass energy set near the threshold for B‑meson pair production (the $\Upsilon$ region). KEKB achieved a peak luminosity of $2.1 \times 10^{34} \, \mathrm{cm}^{-2} \, \mathrm{s}^{-1}$, making it a B‑factory. The primary goals of B factories were precision measurements of B‑meson decays and CP‑violation to test the unitarity of the CKM matrix. Together with the contemporaneous BaBar experiment at SLAC, Belle discovered CP violation in B‑meson decays.
    
In 2010 KEKB/Belle stopped operations for a major upgrade and was transformed into SuperKEKB/Belle II (the next-generation B factory), which began beam commissioning in 2018. SuperKEKB currently runs at a maximum luminosity of $4.7 \times 10^{34} \, \mathrm{cm}^{-2} \, \mathrm{s}^{-1}$ and aims to reach $6 \times 10^{35} \, \mathrm{cm}^{-2} \, \mathrm{s}^{-1}$ by 2040, about 30 times higher than KEKB. In addition to B‑physics, Belle II can cover the entire tau–charm energy region via initial‑state radiation (ISR). Although ISR production cross sections for charmed hadrons are roughly two orders of magnitude smaller, Belle II's \cite{Belle II:Det} ultra‑high luminosity yields effective statistics that will exceed those of BESIII; furthermore, the enormous sample of B‑meson decays at Belle II will also produce abundant charm quarks. 
    
The LHCb experiment \cite{LHCb} at CERN’s Large Hadron Collider, with its unique single-arm forward spectrometer design and excellent vertexing and particle-identification capabilities, is one of the key platforms for heavy-flavor physics and hadron spectroscopy and is often described as a B-hadron factory. Between 2010 and 2018 LHCb collected about 9 $fb^{-1}$ of data and produced a series of major results in heavy-flavor physics and spectroscopy, including observations of CP violation in charm and beauty hadron decays, discoveries of pentaquark and tetraquark states and double-charm baryons, and precision determinations of CKM parameters such as the angle $\gamma$, $\sin 2 \beta$, and $\phi_s$. At $\sqrt{s} = 13$ TeV the charm-hadron production cross section is more than five orders of magnitude higher than at $e^+e^-$ tau–charm machines, giving LHCb vastly larger charm samples. However, the hadronic environment has higher backgrounds and relatively lower detection efficiency for some final states: LHCb is particularly well suited to studies of purely charged-track final states, but is less advantaged for decays involving neutral particles or neutrinos. In 2023 LHCb completed its Upgrade I \cite{LHCbU}, raising the running peak luminosity by a factor of five to $2 \times 10^{33} \, \mathrm{cm}^{-2} \, \mathrm{s}^{-1}$; it is expected to run through 2033 and collect about 50 $fb^{-1}$ of pp collision data. Preparations for Upgrade II, aimed at the HL-LHC era, are underway: Upgrade II plans to increase instantaneous luminosity by a further factor of roughly 5 to 10, with the detector upgrade targeted for completion around 2036 and a goal of accumulating about 300 fb$^{-1}$ of $pp$ data by 2041.
    
At present, the mainstream experimental facilities worldwide engaged in tau-charm–related research—BESIII, LHCb, and Belle II—both compete and complement one another. In summary, in terms of statistical power, BESIII cannot directly compete with LHCb and Belle/Belle II because of the extremely large production cross sections in hadron collisions and the very high luminosities at B factories. However, by leveraging its strengths in threshold production, rich kinematic constraints, and clean backgrounds, BESIII is highly competitive and complementary to LHCb and Belle II in areas such as heavy-flavor hadron spectroscopy, tau-lepton decays, tests of CKM matrix unitarity, and measurements of the internal structure of nucleons and baryons.

\section{Super Tau Charm Facility}
\label{Section4}

\subsection{Accelerator design and luminosity requirements} 
    
Future STCF collider will adopt the large-Piwinski-angle with Crab Waist (CW) scheme \cite{STCF:CrabWaist,STCF:AccOptics}. Overall accelerator construction goals include: build a symmetric double-ring for electrons and positrons with beam energy continuously tunable across 1.0–3.5 GeV per beam (center-of-mass 2.0–7.0 GeV). At a 4 GeV center-of-mass energy the collider luminosity must exceed $0.5 \times 10^{35} \, \mathrm{cm}^{-2} \, \mathrm{s}^{-1}$, with beam lifetimes above 200 s, and maintain reasonable luminosity and lifetime at other physics energy points. Adopt off-axis injection and a constant-current operating mode so that the circulating beam current remains essentially unchanged. The injector must support off-axis injection and deliver high-charge, high-quality electron and positron beams in the 1 to 3.5 GeV range, and should be upgradable to a replacement (swap-out) injection scheme. Globally, the design should retain possibilities for future upgrades (including luminosity increases, extended energy range, and beam polarization at the interaction point).
    
\begin{figure}[tbp]
    \centering
    \includegraphics[width=0.8\textwidth]{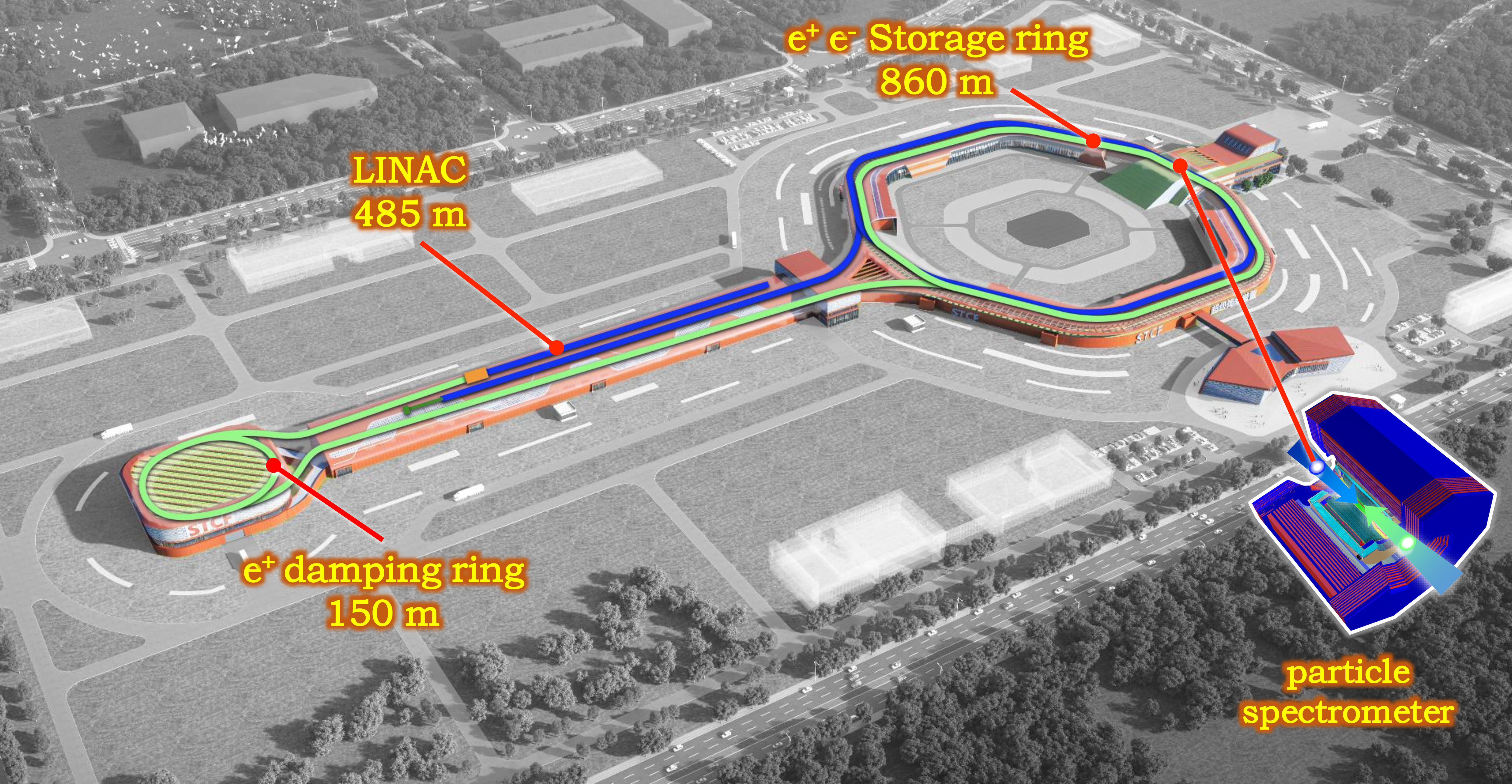}
    \caption{Schematic layout of the Super Tau-Charm Facility}
    \label{fig:STCFlayout}
\end{figure}

The preliminary design is shown in Figure~\ref{fig:STCFlayout}. The storage ring will be a dual-ring, fully symmetric structure of about 860 m in total length, with equal energies for the electron and positron beams. The ring has a single interaction point and two long straight sections designated as the interaction region and the injection region, respectively, with additional straight sections to house damping wigglers and other devices. The linear injector  \cite{STCF:AccInjector} uses full-energy injection, with a total length of about 485 m, and is composed of two electron guns, a beam-focusing system, three linear accelerating sections, a positron source, and a positron damping ring. The injector employs a two-electron-gun scheme: one gun produces bunches that strike the positron-production target to generate positrons \cite{STCF:AccPosSource}, while the other provides electrons for direct injection. This design helps reduce construction costs and is favorable for future polarization and wiggler-based upgrades of the $e^-$ and $e^+$ beams.

The storage ring’s horizontal emittance scales with the square of the energy; higher energy allows a larger single-bunch current, and the threshold for collective effects also increases with energy. Therefore, when the total current is not limited by RF power, the machine’s ultimate luminosity scales roughly as $E^4$. Consequently, in principle, a Super B factory with no technology gap would achieve higher ultimate luminosity than a Super tau-charm factory; at the same luminosity, the Super tau-charm factory would face more severe technical challenges. Early INFN studies have demonstrated the effectiveness of the large-Piwinski-angle plus Crab Waist approach at low luminosity, but at ultra-high luminosity this configuration is currently realized only in the SuperKEKB (the Super B factory), which has not yet reached its design goals. Given this, there is an urgent need for deeper, more detailed studies of the key beam-dynamics and technical challenges for the next generation of high-luminosity circular $e^+e^-$ colliders.

\subsection{Detector technology and requirements for precision measurements}
    
To meet the physics requirements of high-luminosity STCF, we need to design and build a large, integrated particle detector—the STCF detector spectrometer—composed of multiple detector subsystems to detect and identify the final-state particles produced in $e^+e^-$ collisions (protons, pions, kaons, electrons, muons, photons, etc.). A particle-physics experiment demands a detector with as large a solid-angle coverage as possible, high detection efficiency for charged particles, precise momentum measurement and particle identification, and high photon detection efficiency with good energy, position, and time resolutions. Compared with BEPCII (2–5 GeV), the STCF extends the center-of-mass energy range to 2–7 GeV and increases the luminosity by about two orders of magnitude; consequently, the detector design faces challenges of a larger dynamic range, higher radiation backgrounds, higher counting rates, and larger effective event rates. Therefore, the STCF detector design and construction will adopt state-of-the-art technologies, optimized for specific physics goals, with performance expected to meet the physics requirements.
    
\begin{figure}[tbp]
    \centering
    \includegraphics[width=0.8\textwidth]{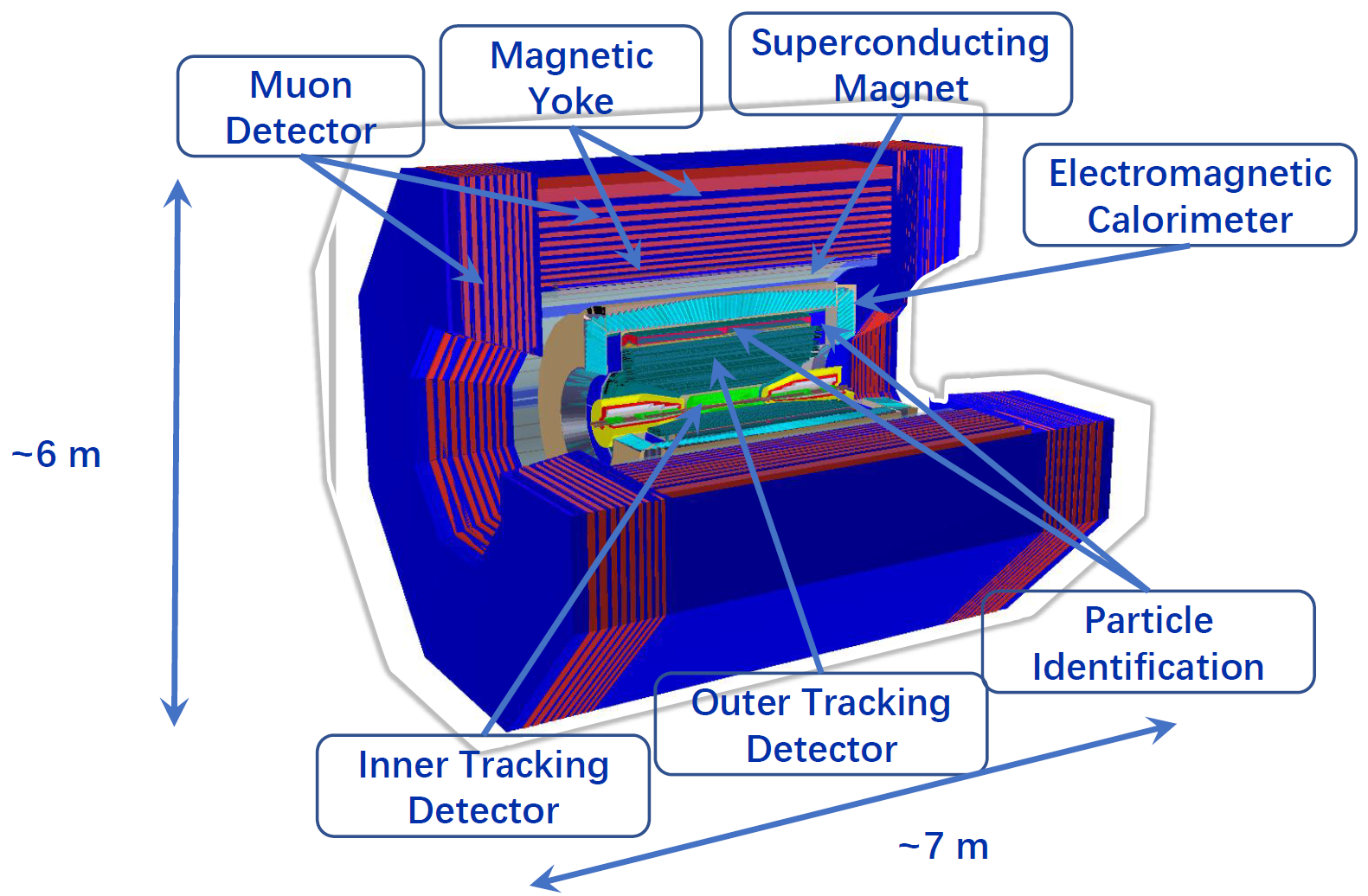}
    \caption{Conceptual design diagram of the STCF spectrometer}
    \label{fig:STCFspectlayout}
\end{figure}

The concept of a STCF detector \cite{STCF:PhyDetCDR} is illustrated in Figure~\ref{fig:STCFspectlayout}. Similar to detectors at other facilities, it comprises from inside out an inner tracking detector, an outer tracking detector, a particle-identification system, an electromagnetic calorimeter, a superconducting magnet, and a muon detector. Since most final-state particles produced at STCF have relatively low momenta, their passage through the detector subjects them to significant material effects \cite{STCF:DetDescr}; thus, in the detector design, the material within the electromagnetic calorimeter and other components should be minimized to optimize low-momentum particle detection. In addition, due to a substantial increase in irradiation doses, the inner-layer detectors must be designed with radiation hardness in mind. 
    
\subsection{Spectrometer: High-precision vertexing, particle ID, and calorimetry}

\textbf{Inner Tracking Detector:} The inner tracker sits immediately inside the beam pipe and is the STCF detector’s innermost layer. Its main role is to work with the central tracking detector to measure charged-particle tracks and momenta and to contribute to vertex reconstruction. Located near the interaction region, it experiences very high radiation, so it must have strong radiation hardness and high-rate capability. The inner tracker plans to use a $\mu$RGroove-based micro-pattern gas detector as the baseline technology. This detector is compact and easy to form into a cylindrical shape; strip readout enables precise position measurements, and a resistive design further reduces material, meeting inner-tracker requirements. The STCF inner tracker will comprise three concentric cylindrical $\mu$RGRoove layers, with a single-layer position resolution of about 100 $\mu m$. To validate the tracking performance of the $\mu$RGroove inner tracker, a proof‑of‑principle prototype had been built and laboratory beam tests has been performed, confirming the feasibility of using the $\mu$RGroove detector as the STCF gaseous inner tracking detector. Another option for the inner tracker detector is the use of a low-power monolithic active pixel sensor, which offers higher spatial resolution and reduced material budget \cite{STCF:Maps,STCF:Maps2}.
       
\textbf{Outer Tracking Detector:} The outer tracker is a large cylindrical precision drift chamber surrounding the inner tracker. It carries the main charged-particle tracking and momentum measurement, and also measures the ionization energy loss ($dE/dx$) for particle identification at low momenta. It is gas-filled and contains many thin wires along the axis to form drift cells. By measuring the drift time of ionization electrons in each cell and the wire positions, the track can be reconstructed. To reduce material, the drift chamber uses a helium-based gas mixture; the inner and outer cylinders are made of carbon fiber; to improve rate capability, a multi-cell nearly square structure shortens drift times. The drift chamber has 48 wire layers, with a single-wire position resolution of about 120 $\mu m$.

\textbf{Particle Identification:} The PID sits close to the outer boundary of the drift chamber and surrounds it, primarily used to distinguish protons, pions, and kaons up to momenta of about 2 GeV/c. To meet this requirement, the STCF PID employs Cherenkov-radiation-based detector technology. The barrel region uses a ring-imaging Cherenkov detector with a liquid C6F14 radiator \cite{STCF:RICH}, while the endcaps use an internally-reflecting Cherenkov detector with a high-purity quartz radiator. Charged particles traversing the radiators emit Cherenkov photons at angles related to their velocity; combining this with the track momentum allows particle identification.
       
\textbf{Electromagnetic Calorimeter:} Surrounding the PID (barrel and endcaps), the EMC measures photon energy and position with high precision under high radiation backgrounds. The STCF EMC uses fast-scintillating pure CsI crystals with large-area APD readout. High-energy photons produce electromagnetic showers in the crystals; by measuring the light yield and the energy distribution, the photon energy and position are determined.
       
 \textbf{Superconducting Magnet:} The superconducting magnet provides a 1 T uniform axial magnetic field. Charged particles bend in the transverse plane, enabling momentum measurement by the tracking detectors. The magnet consists of a superconducting coil and an iron yoke; the yoke serves as the magnetic return path, provides structural support, and acts as an absorber for muons, requiring good mechanical strength.
       
\textbf{Muon Detector:} The muon detector is located in the outermost layer, inserted in the gaps between iron yokes. Since most final-state particles are absorbed by the iron, muons—being highly penetrating—can be detected in these gaps. The STCF muon detector uses a hybrid design of resistive plate chambers (RPC) and plastic scintillators: three inner layers employ bakelite RPCs, and seven outer layers use plastic-scintillator bars \cite{STCF:Muon}. This combination reduces radiation background impact on the muon system while enhancing muon detection efficiency for penetrating neutral and charged particles.

\textbf{Offline software system:} The OSCAR software system  \cite{STCF:SoftFrame} has been developed using advanced tools such as SNiPER, DD4hep, Geant4, ACTS, and AI technologies to meet its specialized needs. OSCAR integrates a complete processing chain covering simulation \cite{STCF:FastSim}, reconstruction, and physics analysis \cite{STCF:CoreSoft}. Key components include: detailed detector modeling with DD4hep \cite{STCF:DetDescr,STCF:OffSoftGeo}; efficient multithreaded simulation and reconstruction; development of high-performance tracking algorithms including a global Hough method and ACTS optimization, achieving over 94\% efficiency for low-momentum tracks; a GNN-based tracking algorithm with strong noise rejection; improved particle identification achieving $4\sigma$ $\pi/K$ separation; and optimized calorimetry performance with $2.15\%$ energy resolution at 1 GeV. Additionally, tools for background mixing, electronics simulation, and advanced machine learning-based identification have been developed and validated. 

\subsection{Beam dynamics, backgrounds, and mitigation strategies} 
Secondary particles produced by intra‑bunch (Touschek) scattering and by beam interactions with the vacuum‑chamber walls or residual gas constitute an important background for the spectrometer, increasing detector noise and radiation load. Studies of beam‑related backgrounds at BEPCII—from both measurements and simulations—identify Touschek scattering as the dominant source. To mitigate this background at BEPCII, a collimation system with movable collimators in the electron and positron rings (horizontal and vertical types) was implemented and studied; horizontal movable collimators played the primary role in these investigations.
    
STCF aims for ultra‑high luminosity at relatively low beam energy, a regime where Touschek effects within electron and positron bunches are enhanced and produce substantially higher beam‑related backgrounds. Consequently, the baseline trigger background during the trigger stage must be higher than for other $e^+e^-$ collider experiments. From extensive detector‑background simulations and characterization studies, the requirement has been set to allow a 5\% false‑positive rate for trigger selection on 1 $\mu s$ data windows — i.e., under nominal background conditions the primary trigger background rate is 50 kHz.
    
Mitigation strategies therefore combine accelerator and detector measures. On the accelerator side these include optimized collimation (with carefully positioned horizontal and vertical movable collimators), improved vacuum and chamber conditioning to reduce beam–gas interactions, beam‑optics tuning to control transverse beam sizes and emittances, and active beam‑stability and feedback systems to limit beam halo and losses. On the detector side, robust shielding, radiation‑hard sensor choices, and fast, background‑aware trigger algorithms are essential.
    
The STCF forward‑region system — situated in the small‑polar‑angle space near the beam pipe — provides real‑time, precise measurements critical for both physics and background control: luminosity and beam‑energy monitoring, detection of key physics signals, and continuous beam‑background surveillance. The detection concept for the three forward subsystems is: a luminosity monitor using diamond detectors combined with Cherenkov detectors for real‑time luminosity measurement; a beam‑energy measurement system built around a laser system and high‑purity germanium (HPGe) detectors for real‑time energy determination; and a zero‑degree event detector using silicon sensors coupled with LYSO crystals for forward/zero‑angle event detection. These instruments enable prompt characterization of background conditions and support active mitigation and interlock strategies for safe, high‑performance operation.

\subsection{Crab Waist collision scheme}
\begin{figure}[htbp]
    \centering
    \includegraphics[width=0.8\textwidth]{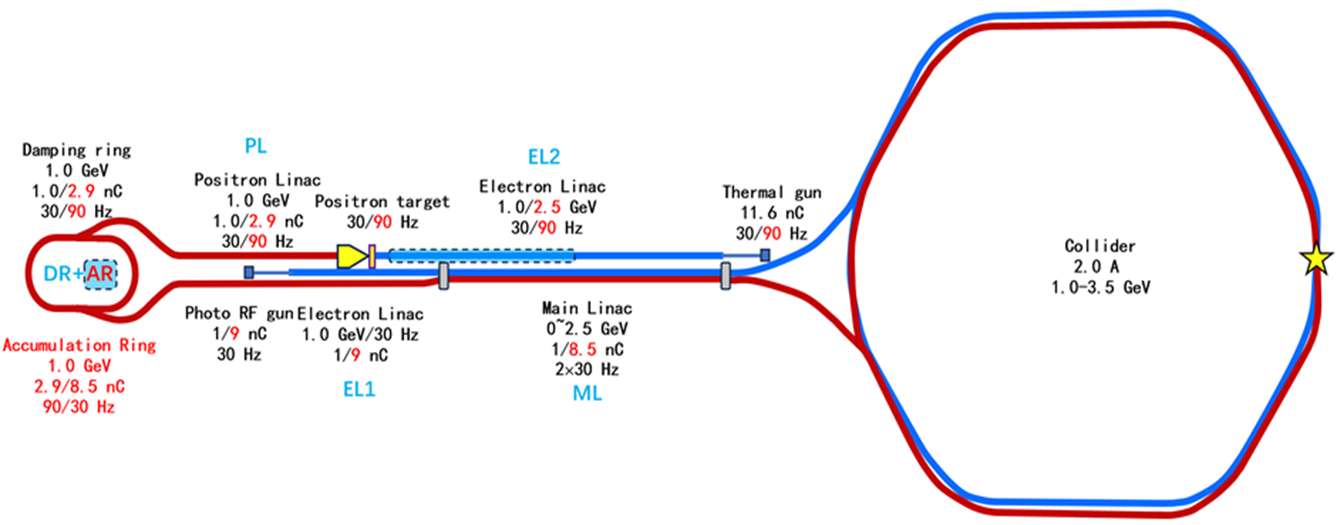}
    \caption{STCF accelerator general layout (Red-highlighted parameters denote the conditions for a potential future upgrade)}
    \label{fig:STCFacclayout}
\end{figure}

To advance ultra-high luminosity, the core technology of the STCF accelerator is to break through peak collision luminosity. STCF will adopt a flat, symmetric electron–positron beam configuration and a double-ring structure, using the large-Piwinski-angle Crab Waist scheme. The corresponding luminosity can be expressed as $\mathscr{L} = \frac{\gamma n_b I_b}{2 e r_e \beta_y^{*} } \, \xi_y H $, where $\gamma$ is the relativistic gamma of electrons or positrons in the lab frame, $n_b$ is the number of bunches, $I_b$ is the per-bunch current, $\xi_y$ is the vertical beam–beam parameter, $H$ is the Hourglass factor representing luminosity loss due to the variation of the envelope along the bunch, $e$ and $r_e$ are the electron charge and classical electron radius, and $\beta_y^{*}$ is the vertical beta at the interaction point. For a collider optimized at a given energy, the luminosity $\mathscr{L}$ scales with the product $n_b$ $I_b$, the vertical beam–beam parameter $\xi_y$, and the Hourglass factor $H$, and is inversely proportional to $\beta_y^{*}$. In 2007, a group at INFN proposed the large crossing angle (large Piwinski angle) with Crab Waist scheme, which was experimentally validated in 2009. The basic idea is to achieve Piwinski angle $\psi \gg 1$ by increasing the crossing angle, reducing natural emittance, and compressing $\beta_x^{*}$. Under this condition, compressing $\beta_y^{*}$ while suppressing the Hourglass effect does not require shortening the bunch length; only $\beta_y^{*} \approx \sigma_x / \theta \ll \sigma_z$ is needed. The large crossing angle also introduces new beam–beam resonances that limit beam–beam interaction parameters; thus, at suitable phases around the interaction point, a pair of crab sextupoles is used to suppress major horizontal coupling resonances and synchro-betatron resonances, further increasing the peak luminosity and expanding the high-luminosity region in phase space. 
    
Future STCF colliders will adopt the large-Piwinski-angle plus Crab Waist scheme; the preliminary design is shown in Figure~\ref{fig:STCFacclayout}. The storage ring will be a dual-ring fully symmetric structure of about 840 m, with equal energies for the electron and positron beams, a single interaction point, two long straight sections for the interaction and injection regions, and additional straight sections to accommodate damping wigglers and other devices. The linear injector uses full-energy injection, with a total length of about 400 m, comprising an electron gun, beam-focusing system, three linear accelerating sections, a positron source, an electron bypass drift, and a positron damping ring. The linear injector employs a single-pulse, two-bunch injection scheme: one bunch targets to produce positrons and the other injects electrons. This design can significantly reduce construction costs and is conducive to future polarization upgrades of the electron and positron beams with wigglers.
 
\subsection{Polarization}

The first‑phase STCF accelerator design will prioritize achieving ultra‑high luminosity and therefore will not include beam polarization in its initial configuration. However, the storage‑ring lattice and civil layout are being developed from the outset to be fully compatible with a future polarization upgrade. In particular, the ring circumference of 860 m has been chosen with reference to international experience to provide ample space for an optimal interaction‑region layout and to simplify integration of long‑length components such as damping wigglers, collimators, injection/extraction systems, and RF cavities.

The ring geometry and straight-section layout preserve space and flexibility for future polarization hardware, including full or partial Siberian snakes, spin rotators, compensating solenoids or helical dipoles, dedicated matching sections, and polarization diagnostics (polarimeters and spin monitors). The lattice supports local optics modifications and matching to integrate these devices with minimal impact on baseline collider optics and performance.

    
The design also considers spin‑dynamics requirements (spin tune control, resonance crossing strategies, spin matching and polarization lifetime) and provides for appropriate routing of services, supports, and cooling for future polarization elements. In this way, STCF can be evolved to deliver polarized beams at the interaction point when scientific priorities and funding permit, while the initial construction remains focused on delivering the required luminosity and detector performance.

\subsection{Expected timeline and milestones}

The indicative project timeline spans 2018–2048. From 2018 to 2020, initial conceptual design work and early CDR preparation were carried out, funded by the Double First Class Initiative at the University of Science and Technology of China (USTC). During 2021–2025 (the 14th Five‑Year Plan), the conceptual design and CDR has been finalized, and key‑technology $\text{R\&D}$ and TDR preparation—supported with CNY 364 million jointly by the Anhui Provincial Government, Hefei Municipal Government, and USTC -- have been executed. A key milestone around the end of 2025 was the submission of the full project proposal to the central government, including the science case, budget, feasibility assessment, and societal impact statement.
    
In 2026–2027 the project will complete the Technical Design Report and civil‑engineering design. Construction, planned to last approximately five years with an estimated cost of CNY 5 billion, is expected to begin in late 2027 or early 2028 and continue through about 2032; this phase will cover civil works, accelerator and detector fabrication, installation, and integration. System integration and pre‑commissioning are anticipated in late 2032–2033, with accelerator and detector commissioning and first beams/data targeted for 2033–2034. Operations are planned to commence in 2034 with an initial running period of 10–15 years (roughly 2034–2044/2049) aiming for $>$ 1 $ab^{-1}$ per year, followed by an upgrade shutdown of about three years and then an additional 10 years of operation. All dates are indicative and remain subject to approval and funding decisions at the 2025 proposal milestone.

\section{Conclusions}
\label{Conclusions}
In summary, the Super Tau--Charm Facility occupies a central position at the precision frontier of high-energy physics. By combining extreme luminosity, threshold operation, and broad energy coverage, STCF addresses foundational questions about how quarks form matter, how fundamental symmetries are realized and violated, and where new physics may first appear in precision observables. It provides a compelling complement to both high-energy colliders and low-energy precision experiments, and its physics reach motivates a broad, sustained experimental program.

\section*{DISCLOSURE STATEMENT}
The authors are not aware of any affiliations, memberships, funding, or financial holdings that might be perceived as affecting the objectivity of this review. 

\section*{ACKNOWLEDGMENTS}
AAP was supported in part by the US Department of Energy grant DE-SC0024357. YZ was supported by the National Key \text{R\&D} Program of China under Contracts No. 2022YFA1602200 and No. 2023YFA1607200; the National Natural Science Foundation of China (NSFC) under Contracts 12221005, 12341501, 12341503, 12341504; the international partnership program of the Chinese Academy of Sciences Grant No. 211134KYSB20200057. We thank the Hefei Comprehensive National Science Center for their strong support on the STCF key technology research project. We thank Haiping Peng, Xingtao Huang, Jianbei Liu, Ye Zou, and Qing Luo for their assistance during the preparation of this manuscript.

%



\end{document}